\documentclass [12 pt]{article}
\def\be{\begin{equation}}
\def\eea{\end{eqnarray}}
\def\bea{\begin{eqnarray}}
\def\ee{\end{equation}}

\def\dis{\hspace{0.1cm}}
\def\pa{\parallel}
\def\fas{\hspace{0.1cm}}
\usepackage[psamsfonts]{amssymb}
\usepackage{amsmath}
\author{Fardin Kheirandish$^{1}$ \footnote{fardin$_{-}$kh@phys.ui.ac.ir} and
Morteza Soltani$^{1}$ \footnote{m.soltani@phys.ui.ac.ir}
\\ $^{1}$ {\small Department of Physics, University of Isfahan,}
\\ {\small Hezar Jarib Ave., Isfahan, Iran.}}
\title{Equivalent approaches to electromagnetic field quantization in a linear
dielectric}
\begin{document}
\maketitle
\begin{abstract}
\noindent It is shown that the minimal coupling method is
equivalent to the Huttner-Barnet and phenomenological approaches
up to a canonical transformation.\\


{\bf PACS number(s): 12.20.Ds}
\end{abstract}
\section{Introduction}
One of the main developments in the quantum optics has been the
study of process, for example spontaneous emission, that take place
in inside, or adjacent to, material bodies. The need to interpret
the experimental results has stimulated attempts to quantize the
electromagnetic field (EM) in a material of general property
\cite{1}.

In an inhomogeneous nondissipative medium, quantization have been
achieved by Glauber and Lewenstein \cite{2}.
 But for dissipative
medium, presence of absorbtion has the effect of coupling the EM
field to a reservoir. This matter is common in different
quantization schemes \cite{3,4,5,6} since in contrast to classical
case, losses in quantum mechanics imply a coupling to a reservoir
whose degrees of freedom have to be added to the Hamiltonian. But in
different methods the coupling between reservoir and EM field has
been considered in different ways.

In damped polarization model the EM field is coupled with matter
field and matter field is coupled with a reservoir. This model is
based on a Lagrangian and equal-time commutation relations (ETCR)
can be written between canonical components. In this approach the
Hamiltonian is obtained from the Lagrangian and is diagonalized by
the Fano technique \cite{3}.

In phenomenological method the reservoir act as a noise source and
these are conveniently represented by the Langevin force which act
on the EM field. The EM field is written in terms of the noise
operators using the Green function of Maxwell equations. The
equivalence of these two methods has been shown in different papers
\cite{4,5,6}.

Recently a new method for dealing with quantum dissipative systems
has been introduced in \cite{7,8} and has been extended to EM field
in a dissipative media. This method is based on a Hamiltonian and
the reservoir couples directly to EM field. In this method the time
revolution of Em field, in contrast to damped polarization model, is
obtained by solving the Heisenberg equations. The extension of this
method to a magnetizable media is straightforward \cite{9}.

The connection between this recent method and the other methods is
the subject of the present paper. Throughout the paper the SI
units are used.

\section{A brief review of different quantization methods}

In this section we review the main approaches to EM field
quantization in a dielectric medium and then compare them.

\subsection{The damped polarization model}

This model is based on the Hopfield model of a dielectric \cite{13}.
The matter is represented by a harmonic polarization field and a
coupling between the polarization field and another harmonic field
is considered. Following the standard approach in quantum
electrodynamics, quantization procedure is started from a Lagrangian
density in real space

\begin{equation}\label{e1}
{\cal L}={\cal L}_{em}+{\cal L}_{mat}+{\cal L}_{res}+{\cal L}_{int},
\end{equation}
where

\begin{equation}\label{e2}
{\cal
L}_{em}=\frac{\epsilon_{0}}{2}\textbf{E}^{2}-\frac{1}{2\mu_{0}}\textbf{B}^{2},
\end{equation}
is the electromagnetic part which can be expressed in terms of the
vector potential $\textbf{A}$ and a scalar potential $U$ ( $
\textbf{B}=\nabla\times \textbf{A},
\textbf{E}=-\dot{\textbf{A}}-\nabla U$). The Lagrangian of the
matter (dielectric) is defined by

\begin{equation}\label{e3}
{\cal
L}_{mat}=\frac{\rho}{2}\dot{\textbf{X}}^2-\frac{\rho\omega_{0}^2}{2}\textbf{X}^2,
\end{equation}
where the field $\textbf{X}$ is the polarization part, modeled by a
harmonic oscillator of frequency $\omega_{0}$. The Lagrangian

\begin{equation}\label{e4}
{\cal L}_{res}=\int_{0}^\infty d\omega
(\frac{\rho}{2}\dot{\textbf{Y}}_\omega^2-\frac{\rho\omega^2}{2}\textbf{Y}_\omega^2),
\end{equation}
is the reservoir part, modeled by a set of harmonic oscillators used
to model the losses. The interaction part

\begin{equation}\label{e5}
{\cal L}_{int}=-\alpha(\textbf{A}\cdot\dot{\textbf{X}}+U\nabla
\cdot\textbf{X})-\int_{0}^\infty d \omega v
(\omega)\textbf{X}\cdot\dot{\textbf{Y}}_{\omega},
\end{equation}
includes the interaction between the light and the polarization
field with coupling constant $\alpha $ and the interaction between
the polarization field and the other oscillators with a frequency
dependent coupling $v(\omega)$. Taking
$\textbf{X}\cdot\dot{\textbf{Y}}_\omega$ as the dissipating term is
not essential but simplifies the calculations. The displacement
field $\textbf{D}(\textbf{r},t)$ is given by the following
combination of electric field and the material polarization

\begin{equation}\label{e6}
 \textbf{D}(\textbf{r},t)=\epsilon_0\textbf{E}(\textbf{r},t)-\alpha
\textbf{X}(\textbf{r},t).
 \end{equation}

 Since $\dot U$ does not appear in the Lagrangian, $U$ is not a proper
 dynamical variable and the Lagrangian can be written in
 terms of the proper dynamical variables $\textbf{A}$, $\textbf{X}$ and
 $\textbf{Y}_\omega$. The easiest way to do this is to go to the
 reciprocal space and write all the fields in terms of spatial
 Fourier transforms. For example the electric field can be written
 as

\begin{equation}\label{e7}
 \textbf{E}(\textbf{r},t)=\frac{1}{(2\pi)^\frac{3}{2}}
\int d^3\textbf{k}\fas\tilde{\textbf{E}}(\textbf{k},t)\fas
e^{i\textbf{k}\cdot\textbf{r}}.
\end{equation}

Since $\textbf{E}(\textbf{r},t)$ is a real field so
$\tilde{\textbf{E}}^*(\textbf{k},t)=\tilde{\textbf{E}}(-\textbf{k},t)$.
Therefore we can restrict integration over $\textbf{k}$ to half
space. The total Lagrangian is

\begin{equation}\label{e8}
L =\int' d^3\textbf{k}(\tilde{{{\cal L}}}_{em}+\tilde{{{\cal
L}}}_{mat}+ \tilde{{{\cal L}}}_{res}+\tilde{{{\cal L}}}_{int}),
\end{equation}
where the prime means that the integration is restricted to the half
of the reciprocal space and the Lagrangian densities in this space
are defined by

\begin{equation}\label{e9}
{\tilde{{\cal L}}_{em}}=\epsilon_0
({\tilde{\textbf{E}}^2}-c^2\tilde{\textbf{B}}^2),
\end{equation}

\begin{equation}\label{e10}
{\tilde{{\cal L}}_{mat}}=\rho
{\dot{\tilde{\textbf{X}}}}^2-\rho\omega_0^2\tilde{\textbf{X}}^2,
\end{equation}

\begin{equation}\label{e11}
\tilde{{\cal L}}_{res}=\int_0^\infty d\omega(\rho
\dot{\tilde{\textbf{Y}}}_\omega^2-\rho\omega^2\tilde{\textbf{Y}}_\omega^2),
\end{equation}

\begin{eqnarray}\label{e12}
\tilde{{\cal L}}_{int}&=&-\alpha[\tilde{\textbf{A}}^*\cdot \dot
{\tilde{\textbf{X}}}+\tilde{\textbf{A}}\cdot \dot
{\tilde{\textbf{X}}}^*+i\textbf{k}\cdot(\tilde{U}^*\tilde{\textbf{X}}-
\tilde{U}\tilde{\textbf{X}}^*)]\nonumber\\&-&\int_0^\infty d\omega
v(\omega)\tilde{\textbf{X}}^*\cdot\dot{\tilde{\textbf{Y}}}_
\omega+\tilde{\textbf{X}}
\cdot\dot{\tilde{\textbf{Y}_\omega^*}}.
\end{eqnarray}

Using the Coulomb gauge $\textbf{k}\cdot
\tilde{\textbf{A}}(\textbf{k},t)=0$ and the Euler-Lagrange equation
for $\dot{\tilde{U}}^*$, we find

\begin{equation}\label{e13}
\tilde{U}(\textbf{k},t)=i\frac{\alpha}{\epsilon_0}(\frac{\textbf{e}_3(\textbf{k})
\cdot{\tilde{\textbf{X}}}(\textbf{k},t)}{k}),
\end{equation}
where $\textbf{e}_3(\textbf{k})$ is the unit vector in the direction
of $\textbf{k}$. The matter fields $\tilde{\textbf{X}}$ and
$\tilde{\textbf{Y}}_\omega$ can be decomposed into transverse and
longitudinal parts. For example $\tilde{\textbf{X}}$ can be written
as

\begin{equation}\label{e14}
\tilde{\textbf{X}}(\textbf{k},t)=\tilde{\textbf{X}}^\pa(\textbf{k},t)
+\tilde{\textbf{X}}^\perp(\textbf{k},t),
\end{equation}
where $\textbf{k}\cdot
\tilde{\textbf{X}}^\perp(\textbf{k},t)=\textbf{k} \times
\tilde{\textbf{X}}^\pa(\textbf{k},t)=0$ and similarly for
$\tilde{\textbf{Y}}_\omega$. The total Lagrangian can then be
written as the sum of two independent parts. The transverse part

\begin{equation}\label{e15}
L^\perp=\int' d^3\textbf{k}({\tilde{\cal
L}_{em}^{\perp}}+{\tilde{{\cal L}}_{mat}^\perp} +{\tilde{{\cal
L}}_{res}^\perp}+{\tilde{{\cal L}}_{int}^\perp})
\end{equation}
where

\begin{equation}\label{e16}
{\tilde{{\cal L} }_{em}^\perp}=\epsilon_0
({\dot{\tilde{\textbf{A}}}^2}-c^2\textbf{k}^2\tilde{\textbf{A}}^2),
\end{equation}

\begin{equation}\label{e17}
{\tilde{{\cal L}}_{mat}^\perp}=(\rho
{\dot{\tilde{\textbf{X}}}}^{\perp2}-\rho\omega_0^2\tilde{\textbf{X}}^{\perp2}),
\end{equation}

\begin{equation}\label{e18}
\tilde{{\cal L}}_{res}^\perp=\int_0^\infty d\omega(\rho
\dot{\tilde{\textbf{Y}}}_\omega^{\perp2}-\rho\omega^2\tilde{\textbf{Y}}_\omega^{\perp2}),
\end{equation}

\begin{equation}\label{e19}
\tilde{{\cal L}}_{int}^\perp=-(\alpha{\tilde{\textbf{A}}}\cdot
\dot{\tilde{\textbf{X}}}^{\perp*}+\int_0^\infty d\omega
v(\omega)\tilde{\textbf{X}}^{\perp*}\cdot\dot{\tilde{\textbf{Y}}}_\omega^\perp+c.c.),
\end{equation}
and the longitudinal part

\begin{equation}\label{e20}
L^\pa =\int' d^3k\tilde{{{\cal L}}}^\pa,
\end{equation}
where

\begin{eqnarray}\label{e21}
\tilde{{\cal L}}^\pa&=&(\rho\dot{\tilde{\textbf{X}}}^{\pa2}-\rho
{\omega_L^2}{\tilde{\textbf{X}}}^{\pa2}+\int_0^\infty
d\omega\rho\dot{\tilde{\textbf{Y}}}_\omega^{\pa2}-\rho\omega^2
\tilde{\textbf{Y}}_\omega^{\pa2})\nonumber\\&-&\int_0^\infty
d\omega(v(\omega)\tilde{\textbf{X}}^{\pa*}\cdot
\tilde{\textbf{Y}}_\omega^{\pa}+c.c.)
\end{eqnarray}

In (\ref{e21}), $\omega_L=\sqrt{\omega_0^2+\omega_c^2}$, is the
longitudinal frequency and
$\omega_c^2=\frac{\alpha^2}{\rho\epsilon_0}$. The link between these
two parts is given by the total electric field

\begin{equation}\label{e22}
\tilde{\textbf{E}}(\textbf{k},t)=\tilde{\textbf{E}}^\perp(\textbf{k},t)+
\tilde{\textbf{E}}^\pa(\textbf{k},t)=
-\dot{\tilde{\textbf{A}}}(\textbf{k},t)+\frac{\alpha}{\epsilon_0}
\tilde{\textbf{X}}^\pa \textbf{e}_3(\textbf{k}).
\end{equation}

Using (\ref{e13}) and the definition of the displacement field
$\textbf{D}$ given in (\ref{e6}), we recover the fact that
$\textbf{D}(\textbf{r},t)$ is a purely transverse field as expected.

The vector potential in reciprocal space can be expanded in terms of
the unit vectors $\textbf{e}_\lambda(\textbf{k})$ as

\begin{equation}\label{e23}
\tilde{\textbf{A}}(\textbf{k},t)=\sum\limits_{\lambda=1,2}
\tilde{A}_\lambda(\textbf{k},t)\textbf{e}_\lambda(\textbf{k}),
\end{equation}
where
$\textbf{e}_\lambda(\textbf{k})\cdot\textbf{e}_\lambda'(\textbf{k})=
\delta_{\lambda\lambda'}$ and
$\textbf{e}_\lambda(\textbf{k})\cdot\textbf{e}(\textbf{k})=0$ for
$\lambda=1,2$.

The Lagrangian ${\cal L}$ can now be used to obtain the components
of conjugate variables

\begin{equation}\label{e24}
-\epsilon_0\tilde{E}_\lambda=\frac{\partial{\cal L}}
{\partial\dot{\tilde{A}}_\lambda^*}=\epsilon_0\dot{\tilde{A}}_\lambda,
\end{equation}

\begin{equation}\label{e25}
\tilde {P}_{\lambda}=\frac{\partial{\cal L}}
{\partial{\dot{\tilde{X}}_\lambda^*}}=
\rho\dot{\tilde{X}}_{\lambda}-\alpha\tilde{A}_\lambda,
\end{equation}

\begin{equation}\label{e26}
\tilde{Q}_{\omega\lambda}= \frac{\partial\pounds}{\partial
\dot{\tilde{Y}}_{\omega\lambda}^*}=\rho\dot
{\tilde{Y}}_{\omega\lambda}-v(\omega)\tilde{Y}_{\omega\lambda}.
\end{equation}

Using the Lagrangian (\ref{e15}) and the expressions for the
conjugate variables (\ref{e24})-(\ref{e26}), we obtain the
transverse part of Hamiltonian as
\begin{equation}\label{e27}
 H^\perp=\int' d^3\textbf{k}(\tilde{\cal H}^\perp_{em}+\tilde{\cal
H}^\perp_{mat}+\tilde{\cal H}^\perp_{int}),
\end{equation}
where

\begin{equation}\label{e28}
\tilde{\cal H}^\perp_{em}=\epsilon_0(\tilde{\textbf{E}})^{\perp2}
+\epsilon_0c^2\tilde{\textbf{k}}^2\tilde{\textbf{A}}^2,
\end{equation}
is the energy density of the EM field, $\tilde{k}$ is defined by
$\tilde{k}=\sqrt{k^2+k_c^2}$ with
$k_c\equiv\frac{\omega_c}{c}=\sqrt{\frac{\alpha^2}{\rho
c^2\epsilon_0}}$, and

\begin{eqnarray}\label{e29}
\tilde{\cal H}^\perp _{mat}  &=&
\frac{\tilde{\textbf{P}}^{\perp2}}{\rho
}\mathop{+\rho\tilde{\omega}_0^2\tilde{\textbf{X}}^{\perp2}+}\int_0^\infty
d\omega (\frac{{\tilde{\textbf{Q}}^\perp_\omega}}{{\rho }}^2+{\rho}
\omega ^2 \tilde{\textbf{Y}}_\omega
^{\perp2})\nonumber\\&+&\int_0^\infty {d\omega }
\frac{{(\nu(\omega)}}
{\rho}\tilde{\textbf{X}}^{\perp*}\cdot\tilde{\textbf{Q}}
^\perp_\omega) + c.c.,
\end{eqnarray}
is the energy density of the matter field which includes the
interaction between the magnetization and the reservoir. The
frequency
 $\tilde{\omega}_0^2\equiv\omega_0^2+\int_0^\infty
d\omega\frac{v(\omega)^2}{\rho^2}$ is the renormalized frequency of
the polarization field and

\begin{equation}\label{e30}
H_{int}^\perp=\frac{\alpha}{\rho}\int{d^3{\bf k}}[{{\bf
A}}^*\cdot\tilde{{\bf P}}+ c.c.],
\end{equation}
is the interaction between the EM field and the magnetization.

By the same method we can obtain the longitudinal part of the
Hamiltonian as

\begin{eqnarray}\label{e31}
\hat{\tilde{\cal H}}^\pa
&=&[{\frac{\hat{\tilde{\textbf{P}}}^{\pa2}}{\rho
}\mathop{+\rho\tilde{\omega}_0^2\hat{\tilde{\textbf{X}}}^{\pa2}+}\int_0^\infty
d\omega (\frac{{\hat{\tilde{\textbf{Q}}}_\omega^\pa}}{{\rho
}}^2+{\rho}\omega^2 \hat{\tilde{\textbf{Y}}}_\omega
^{\pa2})}\nonumber\\&+&{\int_0^\infty {d\omega }
\frac{{(\nu(\omega)}}
{\rho}\hat{\tilde{\textbf{X}}}^{\pa*}\cdot\hat{\tilde{\textbf{Q}}}
_\omega ^\pa) +
c.c.}]+[\frac{\alpha^2}{\epsilon_0}\hat{\tilde{\textbf{X}}}^{\pa2}],
\end{eqnarray}
where we have separated the electric part of the Hamiltonian from
the matter part for later convenience.

The fields are quantized in a standard fashion by demanding ETCR
between the variables and their conjugates. For EM field components
we have

\begin{equation}\label{e32}
[\hat{\tilde{A}}_\lambda(\textbf{k},t),\hat{\tilde{E}}_{\lambda'}^{{*}}
(\textbf{k}',t)]=-i\frac{\hbar}{\epsilon_0}\delta
_{\lambda\lambda'}\delta(\textbf{k}-\textbf{k}'),
\end{equation}
and for the matter

\begin{equation}\label{e33}
[\hat{\tilde{X}}_{\lambda}(\textbf{k},t),\hat{\tilde{P}}_{\lambda'}^{*}
(\textbf{k}',t] =i\hbar \delta_{\lambda,\lambda'}\delta (\textbf{k}-
\textbf{k}'),
\end{equation}

\begin{equation}\label{e34}
[\hat{\tilde{Y}}_{\omega\lambda}(\textbf{k},t),\hat{\tilde{Q}}
_{\omega'\lambda'}^{{*}}(\textbf{k}',t)]=i\hbar\delta
_{\lambda\lambda'}\delta(\textbf{k}-\textbf{k}')\delta(\omega-\omega'),
\end{equation}
with all other equal-time commutators being zero. Indeed, it is
easily shown that the Heisenberg equations of evolution based on the
Hamiltonian (\ref{e27}) and ETCR, (\ref{e32})-(\ref{e34}), are
identical to Maxwell equations in a dielectric medium with
displacement operator $\textbf{D}(\textbf{r},t)$, defined in
(\ref{e6}).

The equations (\ref{e27}) and (\ref{e6}), together with the
commutation relations in (\ref{e32})$-$(\ref{e34}) complete the
quantization procedure of transverse fields.

In order to extract useful information about the system, we need to
solve these equation. To facilitate the calculations, we introduce
two sets of annihilation operators

\begin{equation}\label{e35}
\hat{a}_\lambda(\textbf{k},t) = \sqrt {\frac{\epsilon_0 }{{2\hbar
\tilde{k}c}}}(\tilde{k}c
\hat{\tilde{A}}_\lambda(\textbf{k},t)+i\hat
{\tilde{E}}_{\lambda}(\textbf{k},t)),
\end{equation}

\begin{equation}\label{e36}
\hat{b}_\lambda(\textbf{k},t) = \sqrt {\frac{\rho }{{2\hbar
\tilde{\omega}_0}}}(\tilde{\omega}_0
\hat{\tilde{X}}_\lambda(\textbf{k},t)+\frac{i}{\rho}\hat
{\tilde{P}}_{\lambda}(\textbf{k},t)),
\end{equation}

\begin{equation}\label{e37}
\hat{b}_{\omega\lambda}(\textbf{k},t)=\sqrt{\frac{\rho}{{2\hbar
\omega}}} (-i\omega\hat{\tilde{Y}}_{\omega\lambda}(\textbf{k},t)+
\frac{1}{\rho}\hat{\tilde{Q}}_{\omega\lambda}(\textbf{k},t)),
\end{equation}
where $\tilde{\omega}_0$ is defined in (\ref{e29}). The different
definitions for $\hat{b}$ and $\hat{b}_\omega$ only amount to a
change of phase and have been chosen for further simplicity. From
the ETCR for the fields (\ref{e32})$-$(\ref{e34}) we obtain

\begin{equation}\label{e38}
[\hat{a}_\lambda(\textbf{k},t),\hat{a}_\lambda^\dag(\textbf{k}',t)]=
\delta_{\lambda \lambda'}\delta(\textbf{k}-\textbf{k}'),
\end{equation}

\begin{equation}\label{e39}
[\hat{b}_\lambda(\textbf{k},t),\hat{b}_\lambda^\dag(\textbf{k}',t)]=
\delta_{\lambda \lambda'}\delta(\textbf{k}-\textbf{k}'),
\end{equation}

\begin{equation}\label{e40}
[\hat{b}_{\omega\lambda}(\textbf{k},t),\hat{b}_{\omega'\lambda'}^\dag(
\textbf{k}^\shortmid,t)]=\delta_{\lambda\lambda'}
\delta(\textbf{k}-\textbf{k}')\delta(\omega-\omega').
\end{equation}

We emphasize that, in contrast to the previous ETCR between the
conjugate fields (\ref{e32})-(\ref{e34}), which were correct only in
half $\textbf{k}$ space, the relations (\ref{e38})$-$(\ref{e40}) are
valid in the hole reciprocal space. Inverting eq
(\ref{e35})$-$(\ref{e37}) to express the field operators in terms of
the creation and annihilation operators and inserting them into the
matter part of the Hamiltonian (\ref{e27}), we obtain the normal
ordered Hamiltonian for the transverse part of the matter
Hamiltonian as
\begin{equation}\label{e41}
\hat{H}^\perp=\hat{H}_{em}^\perp+\hat{H}_{matt}^\perp+\hat{H}_{int}^\perp
 \end{equation}
\begin{equation}\label{e42}
\hat{H}_{em}^\perp=\int_0^\infty{d\omega}
  \int{d^3\textbf{k}\sum\limits_{\lambda=1,2}
{\hbar\omega}\hat{a}_\lambda^\dag
  (\omega,\textbf{k})\hat{a}_\lambda(\omega,\textbf{k})},
 \end{equation}

\begin{eqnarray}\label{e43}
\hat{H}_{mat}^\perp&=&\int d^3k\sum\limits_{\lambda=1,2}
[\hbar\tilde{\omega}_0\hat{b}_\lambda^\dagger(\textbf{k},t)
\hat{b}_\lambda(\textbf{k},t)+\int_o^\infty
d\omega\hbar\omega\hat{b}_{\omega\lambda}^\dagger(
\textbf{k},t)\hat{b}_{\omega\lambda}(\textbf{k},t)\nonumber\\
&+&\frac{\hbar}{2}\int_0^\infty
 d\omega
 V(\omega)[\hat{b}^\dagger_\lambda(-\textbf{k},t)+\hat{b}_\lambda
 (\textbf{k},t)][\hat{b}_{\omega\lambda}^\dagger(-\textbf{k},t)
 +\hat{b}_{\omega\lambda}(\textbf{k},t)]],\nonumber\\
\end{eqnarray}

\begin{equation}\label{e44}
\hat{H}_{int}  = i\frac{\hbar}{2}\int {d^3 {\bf
k}}\sum_{\lambda=1,2} \Lambda(k) (\hat a_\lambda (\textbf{k}) + \hat
a_\lambda ^\dag  ( - \textbf{k})) (\hat b_{\lambda } (\textbf{k}) -
\hat b_{\lambda } ^\dag  ( - \textbf{k})) ,
\end{equation}

where
$V(\omega)\equiv[\frac{v(\omega)}{\rho}\sqrt{\frac{\omega}{\tilde{\omega}_0}}]$,
$\Lambda(k)\equiv\sqrt{\frac{\tilde{\omega}_0ck_c^2}{\tilde{k}}}$
and integration over $\textbf{k}$ has been extended to the whole
reciprocal space.

Now instead of solving the Heisenberg equations we use the Fano
technique to diagonalize the Hamiltonian \cite{11}. After
diagonalization the field operators are written in terms of the
eigenoperators of the Hamiltonian.

The diagonalization is down in two step. In the first step the
polarization and reservoir part of the Hamiltonian is digitalized
and the polarization is written in terms of its eigenoperators. Then
the total Hamiltonian is again diagonalized with the same method.

In the first step the diagonalized expression for
$\hat{H}^\perp_{mat}$ is obtained as (the calculations leading to
the diagonalization are lengthy and can be found in \cite{3} and we
do not repeat them here)

\begin{equation}\label{e45}
\hat{H}_{matt}^\perp=\int_0^\infty{d\omega}
  \int{d^3\textbf{k}\sum\limits_{\lambda=1,2}
{\hbar\omega}\hat{B}_\lambda^\dag
  (\omega,\textbf{k})\hat{B}_\lambda(\omega,\textbf{k})},
 \end{equation}
where $\hat{B}^\dagger _\lambda(\textbf{k},\omega)$ and
 $\hat{B}_\lambda(\textbf{k},\omega)$ are the dressed matter field creation
 and annihilation operators and satisfy the usual ETCR

\begin{equation}\label{e46}
[\hat{B}_\lambda(\omega,\textbf{k},t),\hat{B}_{\lambda'}^\dag(\omega',
\textbf{k}',t)]=\delta
_{\lambda\lambda'}\delta(\textbf{k}-\textbf{k}')\delta(\omega-\omega').
\end{equation}

These operators can be expressed in terms of the initial creation
and and annihilation operators as

\begin{equation}\label{e47}
B({\bf k},\omega ) = \alpha _0 (\omega )b({\bf k}) + \beta _0
(\omega )b^\dag  (-{\bf k}) + \int_0^\infty  {d\omega } \alpha
(\omega ,\omega ')b({\bf k},\omega ) + \beta (\omega ,\omega
')b^\dag  (-{\bf k},\omega ),
\end{equation}
and all the coefficient $\alpha_0(\omega)$,
 $\beta_0(\omega)$, $\alpha_1(\omega,\omega')$
 and $\beta_1(\omega,\omega')$ can be obtained in terms of microscopic
 parameters. In \cite{3} the relation between $\alpha_0(\omega)$ and
 $\beta_0(\omega)$ is obtained as
 \begin{equation}\label{e48}
\beta _{0} (\omega) =
\frac{\omega-\tilde{\omega}_0}{\omega+\tilde{\omega}_0
}\alpha_{0}(\omega),
\end{equation}
and we will use this relation in the next section.

Using the commutators of $\hat{b}$ with $\hat{B}$ and $\hat{B}^\dag$
together with (\ref{e47}), we find
\begin{equation}\label{e49}
\hat{b}_\lambda(\textbf{k})=\int_0^\infty{d\omega[\alpha^*_0
(\omega)}\hat{B}_\lambda^\dag(\textbf{k},\omega)-\beta_0(\omega)
\hat{B}_\lambda(-\textbf{k},\omega)].
\end{equation}
The consistency of the diagonalization procedure is checked by
verifying that the initial commutation relation between
$\hat{b}(\textbf{k})$ and $\hat{b}^\dagger(\textbf{k})$ are
conserved.

Using (\ref{e44}), (\ref{e45}), (\ref{e49}) and (\ref{e41}), the
total Hamiltonian  can be written as

\begin{eqnarray}\label{e50}
\hat{H} &=& \int {d^3{\bf k}}\{ \hbar\tilde{\omega}_{\bf k}
\hat{a}_\lambda ^\dag ({\bf k})\hat{a}_\lambda({\bf
k})+\int_0^\infty{d\omega\hbar\omega} \hat{B}_{\lambda }^\dag
({\bf k},\omega )\hat{B}_{\lambda }({\bf k},\omega )\nonumber\\
&+& \frac{\hbar}{2}\Lambda ({k})\int_0^\infty{d\omega}\{ g(\omega
)\hat{B}_{\lambda }^\dag ({\bf k},\omega)[\hat{a}_\lambda({\bf
k})+\hat{a}_\lambda ^\dag(-{\bf k})]+ H.c.\}
\end{eqnarray}
where $\Lambda ({k}) $ is defined in (\ref{e44}) and $g(\omega
)=i\alpha_0(\omega)+i\beta_{0}(\omega)$.

Now we should take the second step and diagonlaize the total
Hamiltonian, (\ref{e50}), as

\begin{equation}\label{e51}
\hat{H}=\int d^3 {\bf k}\int_{0}^\infty\hbar\omega
[\hat{C}^\dag({\bf k}.\omega)\hat{C}({\bf k}.\omega)
\end{equation}
where eigenoperators of the system can be written as
\begin{eqnarray}\label{e52}
\hat{C}({\bf k},\omega ) &=& \tilde{\alpha} _0 (k,\omega
)\hat{a}({\bf k}) + \tilde{\beta} _0 (k,\omega )\hat{a}^\dag
(-{\bf k}) \nonumber\\&+& \int_0^\infty {d\omega' } \tilde{\alpha}
(k,\omega ,\omega ')\hat{B}({\bf k},\omega,\omega' ) +
\tilde{\beta} (k,\omega ,\omega ')\hat{B}^\dag (-{\bf k},\omega'
).
\end{eqnarray}

To calculate the time dependence of EM field operators we should
write them in terms of eigenoperators of the system.

The vector potential is given by

 \begin{equation}\label{e53}
\hat{\textbf{A}}(\textbf{r},t)=\frac{1}{(2\pi)^{\frac{3} {2}}}
\int
d^3\textbf{k}\sum_{\lambda=1,2}\sqrt{\frac{\hbar}{{2\epsilon_0\tilde{\omega}_\textbf{k}}}}
[\hat{a}_\lambda(\textbf{k},t)e^{i\textbf{k}
\cdot\textbf{r}}+H.c.]\textbf{e}_\lambda(\textbf{k}).
\end{equation}

By inverting (\ref{e52}), writing $\hat{a}$ in terms of $\hat{C}$
and using the Hamiltonian (\ref{e51}), to calculate the time
dependence of $\hat{C}$, we obtain $\hat{a}(t)$ as
\begin{equation}\label{e54}
\hat{a}_\lambda({\bf k},t)=\int_0^\infty
{d\omega}[\tilde{\alpha}_0^*({k},\omega)\hat{C}_\lambda({\bf
k},\omega )e^{-i\omega t}-\tilde{\beta} _0 ({k},\omega
)\hat{C}_\lambda^\dag ({\bf k},\omega )e^{i\omega t}].
\end{equation}
Using (\ref{e53}) and (\ref{e54}) we obtain
\begin{eqnarray}\label{e55}
\hat{\textbf{A}}(\textbf{r},t)&=&(\frac{\hbar}{8\pi^4\epsilon_0})^{\frac{1}{2}}\int
{d^3 {\bf k}} \int_0^\infty  {d\omega }
\frac{\omega\sqrt{Im\chi(\omega)}}{{\omega_{\bf k}^2+\omega
^2(1+\chi(\omega))}}\nonumber\\&\times&[\sum\limits_{\lambda=1,2}\hat
C_\lambda({\bf k},\omega )e^{i(\textbf{k}\cdot \textbf{r}-\omega
t)}\textbf{e}_\lambda(\textbf{k}) + H.c.].
\end{eqnarray}
where $\chi(\omega)$ is obtained as
\begin{eqnarray}\label{e56}
\chi(\omega)=\lim_{\varepsilon\longrightarrow0}\frac{1}{2}\int_{-\infty}^{+\infty}{d\omega'}
\frac{{\left|{f(\omega')}\right|^2}}{{\omega'-\omega-
i\varepsilon}}= \frac{1}{2}P\int_{ - \infty }^{ + \infty } {} \{
\frac{{|f(\omega ')|^2 }}{{\omega ' - \omega }}\} d\omega ' +
\frac{1}{2}i\pi
|f(\omega )|^2,\nonumber\\
\end{eqnarray}
and
$f(\omega)=\sqrt{\frac{\alpha^2\omega_0}{\rho\omega^2\epsilon_0}}g(\omega)$.

 The transverse part of EM field can be obtained by the same method using
 the Hamiltonian (\ref{e31}) and the definition of $E^\parallel$ in
 (\ref{e22}) as

\begin{eqnarray}\label{e57}
\hat{ \textbf{E}}^\pa(\textbf{r},t)&=&\left({\frac{\hbar}{8\pi ^4
\epsilon_0}}
  \right)^{\frac{1}{2}}
  \int_0^\infty d\omega\int d^3k\fas\frac{i\sqrt{Im\chi(\omega)}}
 {1+\chi(\omega)}\nonumber\\&\times & (\hat{C}_3(\textbf{k},\omega)e^{i(\omega
 t-\textbf{k}\cdot\textbf{r})}-H.c.)\fas\textbf{e}_3(\textbf{k}).
 \end{eqnarray}

The polarization can be obtained by writing the matter field in
terms of $\hat{C}$ and $\hat{C}^\dagger$ as
\begin{equation}\label{e58}
{\hat{\textbf{P}}}({\bf r},t) = \int_0^\infty  {} d\omega
\{[\epsilon_0\chi (\omega ){\bf \hat{\textbf{E}}}({\bf r},\omega )
+{\hat{\textbf{P}}}_N ({\bf r},\omega )]e^{ - i\omega t}+h.c.\}
\end{equation}
where
\begin{equation}\label{e59}
 \hat{P}_{N\lambda}({\textbf{r}},\omega)=\int d^3 \textbf{k}
 \sqrt{2\hbar\epsilon_0 Im\chi}\hat{C}_{e\lambda}({\bf
 k},\omega)e^{i\textbf{k}.\textbf{r}}.
 \end{equation}
 Also using (\ref{e59}) we can show
that the obtained polarizability satisfies the Kramers-Kronig
relations.

Although the damped polarization model is based on a microscopic
model but as can be seen from (\ref{e55}) and (\ref{e57}), the final
results only depend on the macroscopic parameter $\chi (\omega)$.
Then this model can be used for any medium with known
susceptibility.

The second term of (\ref{e58}) has no classical equivalent and
represents a Langevin fluctuation term (noise operator) which is a
characteristic of a dissipative medium. It is easy to show that this
noise operator satisfy the fluctuation-dissipation theorem
\cite{14}.

\subsection{Phenomenological model}
This model is based on Maxwell equations, Kubo's formula and
dissipation-fluctuation theorem. In accordance to dissipation-
fluctuation theorem, since the medium is dissipative, we should add
to Maxwell equations a noise field. This field is considered as the
source of electromagnetic field. The electromagnetic field can be
written in terms of this noise operators by using the Green function
of classical Maxwell equations.

In this method the Em field operators are separated into the
positive and negative frequency parts in the usual way. For
example

\begin{equation}\label{e60}
\hat{\textbf{E}}(\textbf{r},t)=\frac{1}{\sqrt{2\pi}}\int_0^\infty
d\omega[\hat{\textbf{E}}^+(\textbf{r},\omega)e^{-i\omega
t}+\hat{\textbf{E}}^-(\textbf{r},\omega)e^{i\omega t}],
\end{equation}
where the positive and negative frequency parts involve only the
annihilation and creation operators respectively. Similar
decompositions hold for other field operators. The field operators
satisfy Maxwell equations which in the frequency domain are of the
forms

 \begin{equation}\label{e61}
\nabla\times\hat{\textbf{E}}^+(\textbf{r},\omega)=i\omega\hat{\textbf{B}}^+
(\textbf{r},\omega), \end{equation}

 \begin{equation}\label{e62}
\nabla\times\hat{\textbf{B}}^+(\textbf{r},\omega)=
-i\omega\mu_0\hat{\textbf{D}}^+(\textbf{r},\omega)
\end{equation}
where the monochoromattic electric and displacement field are
related through

\begin{equation}\label{e63}
\hat{\textbf{D}}^+(\textbf{r},\omega)=
\epsilon_0\varepsilon(\omega)\hat{\textbf{E}}^+(\textbf{r},\omega)
+\hat{\textbf{P}}^+_N(\textbf{r},\omega),
\end{equation}
the parameters $\epsilon_0$ and $\mu_0$ are permutability and
permeability of free space respectively and
$\varepsilon(\omega)=1+\chi(\omega)$.

In this equation, $\hat{\textbf{P}}^+_N(\textbf{r},\omega)$ stands
for noise current density operator associated with absorption nature
of the medium. The fluctuation-dissipation theorem and Kubo's
formula require that the noise operator satisfy the following
commutation relation \cite{14}

\begin{eqnarray}\label{e64}
&&
[\hat{P}_{Ni}^{+}(\textbf{r},\omega),\hat{P}_{Nj}^-(\textbf{r}',\omega')]=
2\epsilon_0\hbar Im\chi(\omega)
\delta_{i,j}\delta(\textbf{r}-\textbf{r}')\delta(\omega-\omega'),
\nonumber\\
&&
[\hat{P}_{Ni}^{+}(\textbf{r},\omega),{P}_{Nj}^+(\textbf{r},\omega)]
=[P_{Ni}^-(\textbf{r},\omega),P_{Nj}^{-}(\textbf{r},\omega)]=0.\nonumber\\
\end{eqnarray}

In a gauge in which scaler potential vanishes, we have

\begin{equation}\label{e65}
\hat{\textbf{E}}^+(\textbf{r},\omega)=i\omega\hat{\textbf{A}
}^{+}(\textbf{r},\omega), \end{equation}

 \begin{equation}\label{e66}
\hat{\textbf{B}}^+(\textbf{r},\omega)=
\nabla\times\hat{\textbf{A}}^+(\textbf{r},\omega).\ee

Combining equations (\ref{e62}), (\ref{e63}), (\ref{e65}) and
(\ref{e66}) one can easily show that the positive frequency part of
the vector potential operator satisfies

\begin{equation}\label{e67}
\nabla\times[\nabla\times\hat{\textbf{A}}^+(\textbf{r},\omega)]-\varepsilon
(\omega)
\frac{\omega^2}{c^2}\hat{\textbf{A}}^+(\textbf{r},\omega)=\frac{i\omega}{\epsilon_0c^2}
\hat{\textbf{P}}_N^+(\textbf{r},\omega).
\end{equation}

The differential equation (\ref{e67}) can be converted to an
algebraic equation using a Fourier transformation

 \begin{equation}\label{e68}
-\textbf{k}\times[\textbf{k}\times\hat{\textbf{A}}^+(\textbf{k},\omega)]
-\varepsilon(\omega)\frac{\omega^2}{c^2}\hat{\textbf{A}}^+
(\textbf{k},\omega)=\frac{i\omega}{\epsilon_0c^2}\hat{\textbf{P}}_N^+(\textbf{k},\omega)
\end{equation}

From (\ref{e68}), the positive frequency component of the vector
potential in reciprocal space, can be written as

\begin{equation}\label{e69}
\hat{\textbf{A}}^+(\textbf{k},\omega)=
\frac{i\omega}{\epsilon_0}[{\frac{I-[\frac{c^2}{\omega^2}\varepsilon(\omega)]\textbf{kk}}
{\textbf{k}^2c^2-\omega^2\varepsilon(\omega)}}]\cdot\hat{\textbf{P}}_N^+(\textbf{k},
\omega),
\end{equation}
where $I$ is the unit Cartesian tensor and $\textbf{kk}$ is the
normal Cartesian dyadic. The expression in the bracket is the Green
function of Maxwell equation. We now decompose the Longitudinal and
transverse parts as

\begin{equation}\label{e70}
\hat{\textbf{A}}^{+\perp}(\textbf{k},\omega)=\frac{i\omega}{\epsilon_0}
[{\frac{I-\textbf{e}_3(\textbf{k})\textbf{e}_3(\textbf{k})}
{\textbf{k}^2c^2-\omega^2\varepsilon(\omega)}}]\cdot\hat{\textbf{P}}_N^+(\textbf{k},
\omega),\end{equation} and
 \begin{equation}\label{e71}
\hat{\textbf{A}}^{+\parallel}(\textbf{k},\omega)=\frac{i\omega}{\epsilon_0}
\frac{\textbf{e}_3(\textbf{k})\textbf{e}_3(\textbf{k})}
{{\omega^2\varepsilon(\omega)}}\cdot\hat{\textbf{P}}_N^+(\textbf{k},\omega).
\end{equation}

 Now let us define a new set of boson operators as

\begin{equation}\label{e72}
\hat{{B}}_\lambda(\textbf{k},\omega)= \frac{\hat{{P}}_{N\lambda}^+
(\textbf{k},\omega)}{\sqrt{2\epsilon_0\hbar Im\chi(\omega)}}.
\end{equation}

Using the inverse Fourier transform the operators
$\hat{\textbf{A}}^\perp(\textbf{r},t)$ and
$\hat{\textbf{A}}^\parallel(\textbf{r},t)$ can be obtained from the
equations (\ref{e70}) and (\ref{e71}) as

\begin{eqnarray}\label{e73}
\hat{\textbf{A}}^\perp(\textbf{r},t)&=& ({\frac{\hbar }{{8\pi ^4
\epsilon _0 }}})^{\frac{1}{2}} \int{d\omega\int
{d^3{\textbf{k}}\sum\limits_{\lambda=1,2} \omega \frac{\sqrt
{Im\chi(\omega)}}{\textbf{k}^2c^2-\omega^2
\varepsilon(\omega)}}}\nonumber\\
&\times& (\hat{B}_\lambda(\textbf{k},\omega) e^{i(\omega
t-\textbf{k}\cdot\textbf{r})}+
H.c.)\dis\textbf{e}_\lambda(\textbf{k}),
\end{eqnarray}

\begin{eqnarray}\label{e74}
 \hat{\textbf{A}}^{\pa}(\textbf{r},t)&=& \left({\frac{\hbar}{8\pi^4\epsilon_0}}
 \right)^{\frac{1}{2}}\int_0^\infty{d\omega\int{d^3\textbf{k}\frac{\sqrt
 {Im\chi
 (\omega)}}{\omega\varepsilon(\omega)}}}\nonumber\\
 &\times& (\hat{B}_3(\textbf{k},\omega)e^{i(\omega t-\textbf{k}
 \cdot\textbf{r})}+
 h.c.)\dis\textbf{e}_3(\textbf{k}),
 \end{eqnarray}
and the electric field can be obtained from

\begin{equation}\label{e75}
\hat{\textbf{E}}(\textbf{r},t)=-\frac{\partial\hat{\textbf{A}}(\textbf{r},t)}{\partial
t}.
\end{equation}

The relations (\ref{e73})-(\ref{e74}) are equivalent to the results
of the damped polarization method (\ref{e55}) and (\ref{e57}).
Therefore these two approaches are equivalent.
\subsection{Minimal coupling method}

In the minimal coupling method quantum electrodynamics in a linear
polarizable medium can be accomplished by modeling the medium with a
quantum field, namely a matter field interacting with
electromagnetic field. This quantum field describes the
polarizability character of the medium and interacts with the
displacement field $\hat{\textbf{D}}$ through a minimal coupling
term. \cite{9} (This method is named minimal coupling method since
the matter field is coupled to EM field in a minimal coupling way).
The Heisenberg equations for the electromagnetic field and matter
quantum field lead to both Maxwell and constitutive equations. In
this method we use the Coulomb gauge
$\nabla\cdot{\hat{\textbf{A}}}=0$ and the conjugate canonical
momentum density of electromagnetic field is the displacement vector
operator $\hat{\textbf{D}}(\textbf{r},t)$ which satisfies the
commutation relation

\begin{equation}\label{e76}
 [\hat{\textbf{A}}_\lambda({\textbf{r},t}),\hat{\textbf{D}}_{\lambda'}
 (\textbf{r}',t)]=
 i\hbar\delta_{\lambda\lambda'}^\perp
({\textbf{r}-\textbf{r}'}),
\end{equation}
where $\delta_{\lambda\lambda'}^\perp$ is the transverse delta
function \cite{15}. The total Hamiltonian is written as

\begin{eqnarray}\label{e77}
\hat{H}&=&\int
d^3\textbf{r}({\frac{[\hat{\textbf{D}}(\textbf{r},t)-\hat{\textbf{P}}
(\textbf{r},t)]^2}{2\epsilon_0}}+ {\frac{(\nabla\times
\hat{\textbf{A}}(\textbf{r},t))^2}{2\mu_0}})\nonumber\\
&+&\sum_{\lambda=1,2,3}\int d^3\textbf{k}\int_0^\infty d\omega\hbar
\omega\hat{B}^\dagger_\lambda(\textbf{k},\omega)
\hat{B}_\lambda(\omega,\textbf{k}),
\end{eqnarray}
where $\hat{B}_\lambda(\textbf{k},\omega)$ are matter field
operators which satisfy the boson type commutation relations

\begin{equation}\label{e78}
[\hat{B}_\lambda(\textbf{k},\omega),\hat{B}^{\dag}_{\lambda'}(\textbf{k}',\omega)]=
\delta_{\lambda,\lambda'}\delta(\omega-\omega')\delta(\textbf{k}-\textbf{k}').
\end{equation}

 In the Hamiltonian (\ref{e77}), $\hat{\textbf{P}}(\textbf{r},t)$  is the Polarization
density operator of the medium defined by

\begin{equation}\label{e79}
\hat{\textbf{P}}(\textbf{r},t)=\int
 \frac{d^3\textbf{k}}{(2\pi)^{\frac{3}{2}}}\int_0^\infty{d\omega\sum
 \limits_{\lambda= 1,2,3}
 {[F(\omega)\hat{B}_\lambda
 (\textbf{k},\omega,t)e^{i\textbf{k}.\textbf{r}}+
 H.c.]\textbf{e}_\lambda(\textbf{k})}},
\end{equation}
where the the function $F(\omega)$ is the coupling function
between the electromagnetic field and the matter field.
 Electric field in terms of the displacement vector
${\hat{\textbf{D}}}$ and polarization vector $\hat{\textbf{P}}$
is defined as

\begin{equation}\label{e80}
\epsilon_0\hat{\textbf{E}}(\textbf{r},t)=\hat{\textbf{D}}(\textbf{r},t)-\hat{\textbf{P}}(\textbf{r},t).
\end{equation}
Applying the Heisenberg equation to the vector potential
$\hat{\textbf{A}}$ and using the commutation relations defined in
(\ref{e76}) we find that
\begin{equation}\label{e81}
\hat{\textbf{E}}^\perp(\textbf{r},t)=\frac{\partial\hat{\textbf{A}}(\textbf{r},t)}
{\partial t}\end{equation} and longitudinal part of electric field
is obtained as

\begin{equation}\label{e82}
\hat{\textbf{E}}^\pa(\textbf{r},t)={\hat{\textbf{P}}^\pa(\textbf{r},t)}.
\end{equation}

 Similarly, applying the Heisenberg equation to
the displacement vector $\hat{\textbf{D}}$ we obtain

\begin{equation}\label{e83}
\frac{\partial \hat{\textbf{D}}(\textbf{r},t)}{\partial
t}=\nabla\times \hat{\textbf{H}}(\textbf{r},t).
\end{equation}
Equation (\ref{e80}), (\ref{e81}) and (\ref{e83}) show that the
supposed Hamiltonian (\ref{e77}) gives the correct dynamical
equation (Maxwell equation) and interpreting $\hat{\textbf{P}}$ as
the polarization vector is acceptable.

By using the commutation relation (\ref{e78}), the Heisenberg
equation for $\hat{B}_\lambda
 (\textbf{k},\omega)$ can be obtained as follows

\begin{equation}\label{e84}
 \dot {\hat{B}}_\lambda({\bf k},\omega ,t) =  - i\omega
\hat{B}_\lambda({\bf k},\omega ,t) + \frac{1}{{\hbar \sqrt {(2\pi
)^{\frac{3}{2}}}}}\int_{}^{} {d^3 {\bf r}'\fas F^* (\omega ,{\bf
r}')e^{ - i{\bf k}\cdot{\bf r}'}\hat{\bf E}({\bf r}',t)\cdot{\bf
e}_\lambda ({\bf k})}.
\end{equation}
This equation have the following formal solution
\begin{eqnarray}\label{e85}
\hat{B}_\lambda({\bf k},\omega ,t) &=& \hat{B}_\lambda({\bf
k},\omega )e^{ - i\omega t}\nonumber\\
&+& \frac{1}{{\hbar \sqrt {(2\pi )^{\frac{3}{2}} } }}\int_0^t {}
dt'e^{i\omega (t - t')} \int d{\bf r}'F^* (\omega )e^{ - i{\bf
k}\cdot{\bf r}'}\hat{\bf E}({\bf r}',t')\cdot{\bf e}_\lambda ({\bf
k}).\nonumber\\
\end{eqnarray}
Using this equation and (\ref{e79}), the polarization operator can
be obtained as
\begin{equation}\label{e86}
\hat{\textbf{P}}(\textbf{r},t)=\int_0^{t}
dt'\chi(t-t')\hat{\textbf{E}}(\textbf{r},t')+\hat{\textbf{P}}_N(\textbf{r},t),
\end{equation}
where
\begin{equation}\label{e87}
\chi (t) = \frac{{2 }}{{\hbar \varepsilon _0 }}\int_0^\infty d\omega
|F(\omega )|^2 \sin (\omega t),
\end{equation}
 and $\hat{\textbf{P}}_N(\textbf{r},t)$ is the noise operator
  of the medium and is obtained as
\begin{equation}\label{e88}
\hat{\bf P}_N ({\bf r},t) = \int \frac{d^3{\bf
k}}{(2\pi)^{\frac{3}{2}}}\int_0^\infty d\omega \sum_{\lambda=1}^3
F(\omega )\hat{B}_\lambda({\bf k},\omega )e^{ - i\omega t + i{\bf
k}.{\bf r}}{\bf e}_\lambda({\bf k})+H.c.
\end{equation}
From (\ref{e87}), the imaginary and real part of the Fourier
transform of susceptibility can be written as

\begin{equation}\label{e89} Im
[\chi(\omega)]=\frac{{\pi}}{\hbar\epsilon_0}
|F(\omega)|^2,\end{equation}

 \begin{equation}\label{e90}
 Re[\chi(\omega)]=\frac{\pi}{\hbar\epsilon_0}P\int_{-\infty}^{+\infty}
d\omega|F (\omega')|^2 \frac{\omega}{\omega^2-\omega'^2}.
\end{equation}
As can be seen from (\ref{e89}) and (\ref{e90}), $Im\chi(\omega)$
and $Re\chi(\omega)$ satisfy the Kramers-Kronig relations, which
verifies the interpreting of $\chi(t)$ as the susceptibility.

In (\ref{e88}), we can separate the positive and negative components
of the noise operator $\hat{\bf P}_N ({\bf r},t)$ as defined in the
phenomenological method and then using (\ref{e89}) and (\ref{e78})
we find
\begin{equation}\label{e91}
[\hat{P}_{Ni}(\textbf{r},\omega),\hat{P}_{Nj}(\textbf{r}',\omega')]=
2\epsilon_0\hbar Im\chi(\omega)\delta_{i
j}\delta(\omega-\omega')\delta(\textbf{r}-\textbf{r}'),
\end{equation}
which is the pustulated relation in the phenomenological method
(\ref{e64}).

In a physical situation where the electric susceptibility is known,
the coupling function can be obtained from (\ref{e89}) and the
Hamiltonian (\ref{e77}) can be constructed in terms of the electric
susceptibility. Then the equation of motion can be obtained from
(\ref{e80}), (\ref{e81}) and (\ref{e83}) as

\begin{eqnarray}\label{e92}
-\nabla^2\hat{\textbf{A}}&+&\frac{1}{c^2}\frac{\partial^2\hat{\textbf{A}}}{\partial
t^2 }+\frac{1}{c^2}\frac{\partial}{\partial
t}\int_0^tdt'\chi(t-t')\frac{\partial
\hat{\textbf{A}}(\textbf{r},t')}{\partial
t'}\nonumber\\&=&\mu_0\frac{\partial
\hat{\textbf{P}}_N^\perp(\textbf{r},t)}{\partial t}.
\end{eqnarray}

In this equation if we change the lower limit of integral from $0$
to $-\infty$ and take the time-Fourier transform of this equation we
can obtain the transverse part of the postulated equation of the
phenomenological method (\ref{e68}). Also the time-Fourier transform
of the longitudinal part of electric filed, $E^\pa$, is equivalent
to the longitudinal part of the postulated equation of the
phenomenological method. Then as can be seen the phenomenological
method can be obtained from the minimal coupling method.

Equation (\ref{e92}) can be solved by going to the reciprocal space.
The vector potential in reciprocal space can be written as

\begin{equation}\label{e93}
\hat{\textbf{A}}(\textbf{r},t)=\frac{1}{(2\pi)^{\frac{3}{2}}}\int
d^3\textbf{k}\hat{\tilde{\textbf{A}}}(\textbf{k},t)e^{i\textbf{k}\cdot\textbf{r}},
\end{equation}
 where the Fourier component $\hat{\tilde{\textbf{A}}}$ can be written in
terms of creation and annihilation operators

\begin{equation}\label{e94}
\hat{\tilde{\textbf{A}}}(\textbf{k},t)=\sum_{\lambda=1,2}\sqrt{\frac{\hbar}
{2\epsilon_0\omega_\textbf{k}}} [\hat{a}_\lambda(\textbf{k},t)
\textbf{e}_\lambda(\textbf{k})+\hat{a}^\dagger_\lambda(-\textbf{k},t)
\textbf{e}_\lambda(-\textbf{k})].
\end{equation}
In terms of the Fourier components, equation (\ref{e94}) is written
as

\begin{eqnarray}\label{e95}
\ddot{\hat{\tilde{\textbf{A}}}}&+&\omega_\textbf{k}^2\hat{\tilde{\textbf{A}}}+
\frac{\partial}{\partial
t}\int_0^tdt'\chi(t-t')\dot{\hat{\tilde{\textbf{A}}}}(\textbf{k},t')\nonumber\\&=&-
\frac{1}{\epsilon_0}
\sum_{\lambda=1,2}\int_0^\infty{d\omega}[\omega F(\omega)
\hat{B}_\lambda(\textbf{k},\omega)e^{-i\omega t}\textbf{e}_\lambda
(\textbf{k})+H.c.], \nonumber\\
\end{eqnarray}
where $\omega_\textbf{k}=c|\textbf{k}|$.

The equation (\ref{e95}) can be solved by using the Laplace
transformation, the details can be found in reference \cite{9},
and the final result is

\begin{eqnarray}\label{e96}
\hat{\textbf{A}}(\textbf{r},t)&=&\int{d^3\textbf{k}}
\sum\limits_{\lambda=1,2}{\sqrt{\frac{\hbar}
{2(2\pi)^3\epsilon_0\omega_\textbf{k}}}
[z(\omega_\textbf{k},t)e^{i\textbf{k}\cdot\textbf{r}}
\hat{a}_\lambda{(\textbf{k}},0)}+H.c.]\textbf{e}_\lambda(\textbf{k})\nonumber \\
&+& \frac{1}{\epsilon_0}\sum\limits_{\lambda=1,2}
{\int\frac{d^3\textbf{k}}{(2\pi)^{\frac{3}{2}}}
 \int_0^\infty{d\omega}[\xi(\omega,\omega_\textbf{k},t)
 \hat{B}_\lambda(\textbf{k},\omega,0)}e^{i\textbf{k}\cdot\textbf{r}}+ H.c.]
 \textbf{e}_\lambda(\textbf{k}),\nonumber\\
\end{eqnarray}
where

\begin{equation}\label{e97}
z(\omega _{\textbf{k}},t)=L^{-1}\{\frac{[s+s\tilde\chi(s)-
i\omega_{\textbf{k}}]}{s^2+\omega_\textbf{k}^2+s^2\tilde\chi(s)}\},
\end{equation}

\begin{equation}\label{e98}
\xi(\omega,\omega_\textbf{k},t)=F(\omega)L^{-1}\{
\frac{s}{(s+i\omega)[s^2+ \omega_\textbf{k} ^2  + s^2
\tilde{\chi}(s)]}\},
\end{equation}
and $L^{-1}\{.\}$ denotes the inverse Laplace transform. The
transverse electric field is obtained as

\begin{equation}\label{e99}
\hat{\textbf{E}}^\perp(\textbf{r},t)=-\frac{\partial\hat{\textbf{A}}(\textbf{r},t)}{\partial
t},
\end{equation}
and

\begin{equation}\label{e100}
\hat{\textbf{B}}(\textbf{r},t)=\nabla\times
\hat{\textbf{A}}(\textbf{r},t).
\end{equation}

From equations (\ref{e82}) and (\ref{e85}) the longitudinal part of
electric field is

\begin{eqnarray}\label{e101}
&&\hat{\textbf{E}}^{\pa}(\textbf{r},t)=-\frac{\hat{\textbf{P}}^{\pa}(\textbf{r},t)}{{\varepsilon
_0 }}\nonumber\\&=&-\frac{1}{\varepsilon_0}{\int_0^\infty{d\omega }
\int {d^3\textbf{k}}[Q(\omega
,t)}F(\omega)\hat{B}_3(\textbf{k},\omega,0)e^{i\textbf{k}\cdot
\textbf{r}}
+ H.c.]\textbf{e}_3(\textbf{k}),\nonumber\\
\end{eqnarray}
where

\begin{equation}\label{e102}
Q(\omega, t)=L^{-1}\{ \frac{1}{(s+i\omega)(1+\tilde\chi(s))}\}.
\end{equation}
In section 4 we will show that these results,
(\ref{e96})$-$(\ref{e102}), are equivalent to the results obtained
in the previous methods.
\section{Derivation of the minimal coupling Hamiltonian from the
 damped polarization model}

In this section we want to obtain the Hamiltonian of minimal couplig
method from the Lagrangian of damped polarization model. In
(\ref{e32}) the conjugate of $\tilde{\textbf{A}}$ is the transverse
electric field $-\epsilon_0\tilde{\textbf{E}}^\perp$ but in minimal
coupling scheme, as given in (\ref{e76}), the conjugate of
 $\tilde{\textbf{A}}$ is the displacement vector
 $\tilde{\textbf{D}}$. We can  change the conjugate of $\hat{\textbf{A}}$ from $\hat{\textbf{E}}$
 to $\hat{\textbf{D}}$ by adding to the Lagrangian (\ref{e1}) the term
 $\alpha\frac{\partial (\textbf{X}\cdot \textbf{A})}{\partial t}$ which is
 a canonical transformation and do not effect the equations of motion.
 This canonical transformation leads to a
 $\dot{\bf A}\cdot\textbf{X}$ type of coupling which gives the
displacement field $-\tilde{\textbf{D}}$ as the conjugate of
$\tilde{\textbf{A}}$. In this case using the Fourier transform the
transverse part of the canonically transformed Lagrangian
($\tilde{{\cal L}}'^\perp$) can be written as

\begin{equation}\label{e103}
\tilde{{\cal L}}'^\perp= ({\tilde{{\cal
L}}_{em}^\perp}+{\tilde{{\cal L}}_{mat}^\perp} +{\tilde{\cal
L}_{res}^\perp}+\tilde{{\cal L}}_{int}'^\perp),
\end{equation}
where

\begin{equation}\label{e104}
\tilde{{\cal L}}_{int}'^\perp=(\alpha\dot{\tilde{\textbf{A}}}\cdot
\tilde{\textbf{X}}^{\perp*}+c.c.)-(\int_0^\infty d\omega
v(\omega)\tilde{\textbf{X}}^{\perp*}\cdot\dot{\tilde{\textbf{Y}}}^\perp+c.c.).
\end{equation}

The other transverse or longitudinal parts of the Lagrangian do not
change. Now ${\cal L}'$ is used to obtain the components of the
conjugate variables of the fields

\begin{equation}\label{e105}
-\tilde{D}_\lambda=\frac{\partial{\cal
L}}{\partial\dot{\tilde{A}}_\lambda^*}
=\epsilon_0\dot{\tilde{A}}_\lambda+\alpha\tilde{X}_\lambda^\perp,
\end{equation}

\begin{equation}\label{e106}
\tilde {P}_\lambda^{\perp}=\frac{\partial{\cal L}}
{\partial{\dot{\tilde{X}}^{\perp*}_\lambda}}=
\rho\dot{\tilde{X}}^{\perp}_{\lambda},
\end{equation}

\begin{equation}\label{e107}
\tilde{Q}_{\omega\lambda}^{\perp}= \frac{\partial{\cal L}}{\partial
\dot{\tilde{Y}}_{\omega\lambda}^{\perp*}}=\rho\dot
{\tilde{Y}}_{\omega\lambda}^{\perp}-v(\omega)\tilde{Y}_\lambda^{\perp},
\end{equation}
where (\ref{e104}) is used to obtaining (\ref{e105})$-$(\ref{e107}).
Using the Lagrangian (\ref{e103}) and the expression for the
conjugate variables in (\ref{e105})$-$(\ref{e107}) we obtain the
Hamiltonian for the transverse field as

\begin{equation}\label{e108}
H^\perp=\int' d^3\textbf{k}(\tilde{\cal H}_{em}^\perp+\tilde{\cal
H}_{mat}^\perp),
\end{equation}
where

\begin{equation}\label{e109}
\tilde{\cal H}_{em}^\perp=(\frac{\tilde{\textbf{D}}
+\alpha\tilde{\textbf{X}}^\perp}{\epsilon_0})^2 +\frac
{\textbf{k}^2\tilde{\textbf{A}}^2}{\mu_0},
\end{equation}
is the electromagnetic energy density. The interaction between
electromagnetic field and the polarization field is embodied in
electromagnetic energy density. Using (\ref{e6}) we can rewrite
(\ref{e109}) as

\begin{equation}\label{e110}
\tilde{\cal H}_{em}^\perp=\epsilon_0\tilde{\textbf{E}}^2(\textbf{k})
+\epsilon_0c^2\tilde{\textbf{B}}^2(\textbf{k}).
\end{equation}
This Hamiltonian is like the Hamiltonian of the free EM field. But
since in this Hamiltonian electric field
$\tilde{\textbf{E}}(\textbf{k})$ is not the conjugate component of
vector potential $\tilde{\textbf{A}}(\textbf{k})$ then we can not
use it to separate the eagenmode of the Hamiltonian and quantizing
the EM field.

Again the fields are quantized by demanding the ETCR between
components and its conjugates. Then we should change the equation
(\ref{e24})  as
\begin{equation}\label{e111}
[\hat{\tilde{A}}_\lambda(\textbf{k},t),\hat{\tilde{D}}_{\lambda'}^{{*}}
(\textbf{k}',t)]=i\hbar\delta
_{\lambda\lambda'}\delta(\textbf{k}-\textbf{k}').
\end{equation}

 The matter and reservoir parts of the Hamiltonian
 do not change and we can use the result of
section 2-1 to write
\begin{equation}\label{e112}
\hat{H}_{matt}^\perp=\int_0^\infty{d\omega}
  \int{d^3\textbf{k}\sum\limits_{\lambda=1,2}
{\hbar\omega}\hat{B}_\lambda^\dag
  (\textbf{k},\omega)\hat{B}_\lambda(\textbf{k},\omega)},
 \end{equation}
where $\hat{B}_\lambda
  (\textbf{k},\omega)$ and $\hat{B}_\lambda^\dag
  (\textbf{k},\omega)$ are defined in (\ref{e47}).

 The electromagnetic part of the Hamiltonian has
been written in reciprocal space and can be inverted to real space
by using the inverse Fourier transform as

\begin{equation}\label{e113}
\hat{H}_{em}^\perp =\int{d^3{\bf r
}\frac{(\hat{\textbf{D}}(\textbf{r},t)+\alpha\hat{\textbf{X}}^\perp
(\textbf{r},t))^2 }{2\varepsilon_0}+\frac{(\nabla\times
{\hat{\textbf{A}}(\textbf{r},t)})^2}{2\mu _0}},
\end{equation}
where $\hat{\textbf{X}}^\perp(\textbf{r},t) $ can be written in
terms of its Fourier component
$\hat{\tilde{\textbf{X}}}^\perp(\textbf{k},t) $ as

\begin{equation}\label{e114}
\hat{\textbf{X}}^\perp(\textbf{r},t)=\frac{1}{(2\pi)^\frac{3}{2}}
\int d^3{\bf k}\fas\hat{\tilde{\textbf{X}}}^\perp(\textbf{k},t)\fas
e^{i\textbf{k}\cdot\textbf{r}}.
\end{equation}

Now using the definition of $\hat{b}_\lambda(\textbf{k},t)$ and
$\hat{b}_\lambda^\dagger(\textbf{k},t)$ in (\ref{e36}), we obtain
$\hat{\textbf{X}}^\perp(\textbf{r},t) $ as

\begin{equation}\label{e115}
\hat{\textbf{X}}^\perp(\textbf{r},t)=\frac{1}{(2\pi)^{\frac{3}{2}}}
\sqrt{\frac{\hbar}{{2\rho\tilde{\omega}_0}}} \int
d^3\textbf{k}\sum_{\lambda=1,2}[\hat{b}_\lambda(\textbf{k},t)
e^{i\textbf{k}\cdot\textbf{r}}+h.c.]\fas\textbf{e}_\lambda(\textbf{k}).
\end{equation}

Using (\ref{e49}) and (\ref{e115})we have
 \begin{equation}\label{e116}
 \hat{\textbf{X}}^\perp(\textbf{r},t)=-\frac{1}{(2\pi)^{\frac{3}{2}}}\int_0^\omega
{d\omega}\int{d^3 \textbf{k}\sum\limits_{\lambda= 1,2}[
{\frac{F(\omega)}{\alpha}}\hat{B}_\lambda
(\textbf{k},\omega,t)e^{i\textbf{k}.\textbf{r}}+
H.c.]\textbf{e}_\lambda(\textbf{k})}.
\end{equation}
where
\begin{equation}\label{e117}
F(\omega)=-\sqrt{\frac{\hbar\alpha^2}{2\rho\tilde{\omega}_0}}(
{\alpha_0^{*}(\omega)-\beta_0^{*}(\omega)}).
\end{equation}

 Now using (\ref{e116}) and (\ref{e113}) the total transverse
Hamiltonian (\ref{e108}) can be written as

\begin{eqnarray}\label{e118}
 \hat{H}^\perp&=&\hat{H}_{em}^\perp+\hat{H}_{mat}^\perp\nonumber\\
 &=&\int{d^3\textbf{r}}
 \frac{{\left({\hat{\textbf{D}}-\hat{\textbf{P}}^\perp}
 \right)^2}}
 {{2\varepsilon_0}}
+\frac{(\nabla \times {\hat{\textbf{A}}})^2}{2\mu_0}\nonumber\\&+&
 \int_0^\omega{d\omega}
  \int{d^3\textbf{k}\sum\limits_{\lambda=1,2}
{\hbar\omega}\hat{B}_\lambda^\dag
  (\textbf{k},\omega)\hat{B}_\lambda(\textbf{k},\omega,t)},
 \end{eqnarray}
where

\begin{equation}\label{e119}
\hat{\textbf{P}}^\perp(\textbf{r},t)=\frac{1}{(2\pi)^{\frac{3}{2}}}\int_0^\omega{d\omega}
 \int{d^3\textbf{k}\sum\limits_{\lambda=1,2}[F(\omega)\hat{B}_\lambda
 (\textbf{k},\omega,t)e^{i\textbf{k}\cdot\textbf{r}}+H.c.]\textbf{e}_\lambda(\textbf{k})}.
 \end{equation}

Using the same method the Longitudinal part of the matter field can
be written as
\begin{equation}\label{e120}
 \hat{\textbf{X}}^\pa(\textbf{r},t)=\frac{1}{(2\pi)^{\frac{3}{2}}}\int_0^\infty
{d\omega}\int{d^3\textbf{k}[\frac{F(\omega)}{\alpha}\hat{B}_3
(\textbf{k},\omega,t)e^{i\textbf{k}\cdot\textbf{r}}+
h.c.]}\fas\textbf{e}_3(\textbf{k}).
\end{equation}

The total longitudinal part of the Hamiltonian (\ref{e31}) can be
written as

\begin{eqnarray}\label{e121}
 \hat{H}^\pa&=&\hat{H}_{em}^\pa+\hat{H}_{mat}^\pa\nonumber\\
 &=&\frac{1}{(2\pi)^{\frac{3}{2}}}
 \int{d^3\textbf{r}}\frac{{\left({\int_0^\infty{d\omega}
 \int{d^3\textbf{k}[F(\omega)\hat{B}_3
(\textbf{k},\omega)e^{i\textbf{k}\cdot\textbf{r}}+H.c.]\textbf{e}_3(\textbf{k})}}
 \right)^2}}{{2\varepsilon_0}}
 \nonumber\\&+&\int_0^\omega{d\omega}
  \int d^3\textbf{k}\fas
{\hbar\omega}\fas{B}_3^\dag
  (\textbf{k},\omega)\hat{B}_3(\textbf{k},\omega),
 \end{eqnarray}
where $F(\omega)$ is defined in (\ref{e117}). Finally the total
Hamiltonian can be written as the sum of transverse part
(\ref{e118}) and longitudinal part (\ref{e121})

\begin{eqnarray}\label{e122}
 \hat{H} &=& \hat{H}^\perp + \hat{H}^{\parallel}\nonumber\\
 &=&\int {d^3\textbf{r}}
 \frac{{\left( {\hat{\textbf{D}}-\hat{\textbf{P}}}\right)^2}}
 {{2\varepsilon _0}}+\frac{{(\nabla  \times {\hat{\textbf{A}}})^2 }}
 {{2\mu _0}}
 \nonumber\\&+&\int_0^\omega{d\omega}\int{d^3\textbf{k}\sum
 \limits_{\lambda=1,2,3}{\hbar\omega }
 \hat{B}_\lambda^\dag(\textbf{k},\omega)\hat{B}_\lambda(\textbf{k},\omega)}.\nonumber\\
 \end{eqnarray}
where

 \begin{equation}\label{e123}
 \hat{\textbf{P}}=-\frac{1}{(2\pi)^{\frac{3}{2}}}
 \int_0^\omega  {d\omega }
 \int {d^3\textbf{k}\sum\limits_{\lambda= 1,2,3} [ F(\omega)
 \hat{B}_\lambda
 (\textbf{k},\omega)e^{i\textbf{k}\cdot\textbf{r}}+H.c.]\fas
 \textbf{e}_\lambda(\textbf{k})}
 \end{equation}

Therefore the minimal coupling Hamiltonian is obtained from the
Lagrangian (\ref{e1}) by using a canonical transformation then the
two Hamiltonians in relation (\ref{e27})$-$(\ref{e30}) and
(\ref{e77}) are equivalent.

We can check the consistency of minimal coupling method with damped
polarization method by comparting the obtained susceptibility from
these two methods. In the damped polarization model the imaginary
part of the susceptibility $Im\chi(\omega)$ is obtained as

\begin{equation}\label{1}
Im\chi(\omega)={\frac{\alpha^2\omega_0\pi}{2\rho\omega^2\epsilon_0}}
(\alpha(\omega)+\beta(\omega))^2,
\end{equation}
where we have used Eq. (\ref{e56}) and the definition of $g(\omega)$
in (\ref{e50}). In minimal coupling method $Im\chi(\omega)$ can be
written as

\begin{equation}\label{2}
Im\chi(\omega)={\frac{\pi\alpha^2}{2\rho\omega_0\epsilon_0}}
(\alpha(\omega)-\beta(\omega))^2,
\end{equation}
where we have used (\ref{e89}) and (\ref{e117}). Now from
(\ref{e48}) we can easily prove

\begin{equation}\label{3}
(\alpha(\omega)-\beta(\omega))^2=\frac{\omega_0^2}{\omega^2}
(\alpha(\omega)+\beta(\omega))^2,
\end{equation}
Substituting ({\ref{3}) in (\ref{2}), relation (\ref{1}) will be
obtained.

\section{Comparing the results of the minimal coupling method with other methods}

In this section we want to complete the equivalence of the minimal
coupling method with other methods and show that the results
obtained in the minimal coupling method, (\ref{e96})$-$(\ref{e102}),
are equivalent to those obtained in other methods. In equations
(\ref{e96})$-$(\ref{e102}) the vector potential is obtained as a
combination of the free field operators and the reservoir field
operator respectively as

\begin{equation}\label{e124}
\hat{\textbf{A}}(\textbf{r},t)=I+II,
\end{equation}
where by setting the initial time at $-\infty$, we can write the
expressions for $I$ and $II$ as

\begin{eqnarray}\label{e125}
I=\mathop{\lim}\limits_{t'\longrightarrow-\infty}\int{d^3\textbf{k}}
\sum\limits_{\lambda=1,2}{\sqrt{\frac{\hbar}{2(2\pi)^3
\varepsilon_0\omega_\textbf{k}
}}}[z_+(\omega_{\textbf{k}},t-t')e^{i\textbf{k}\cdot\textbf{r}}\hat{a}_\lambda
(\textbf{k},t')+
H.c.]\textbf{e}_\lambda(\textbf{k}),\nonumber\\
\end{eqnarray}
and

\begin{equation}\label{e126}
II=\lim\limits_{t'\to-\infty}
\frac{1}{\varepsilon_0}\sum\limits_{\lambda=1,2}{\int{\frac{d^3
\textbf{k}}{(2\pi)^{\frac{3}{2}}}}\int_0^\infty{\omega}[\xi(\omega}
,\omega_\textbf{k},t)\hat{B}_\lambda(\textbf{k},\omega,t')
e^{i\textbf{k}\cdot\textbf{r}}+H.c.]\textbf{e}_\lambda(\textbf{k}).
\end{equation}

In other words in Eqs.(\ref{e96}) and (\ref{e101}) we change time
variable from $t$ to $t-t'$ and then let $t'\rightarrow -\infty$.

The explicit time dependence of $z(\omega_\textbf{k},t-t')$ and
$\xi(\omega_\textbf{k},\omega,t-t')$ can be obtained from the the
inverse Laplace transform formula

\begin{equation}\label{e127}
f(t)=L^{-1}[f(s)]=\mathop{\lim}\limits_{\eta\to 0}\int_{-
\infty}^{+\infty}{\frac{f(i\omega
+\eta)}{2\pi}}e^{(i\omega+\eta)t}d\omega,
\end{equation}
where $f(t)$ is an arbitrary function. We have

\begin{eqnarray}\label{e128}
&&\lim\limits_{t'\to-\infty}z_+(\omega_\textbf{k},t-t')
=\nonumber\\&&\lim\limits_{\eta\to 0^+}\lim \limits_{t'\to-
\infty}\int_{-\infty}^{+\infty}
{\frac{d\omega}{2\pi}\frac{(i\omega+\eta) [1+\chi
(-\omega)]-i\omega_\textbf{k}}{\omega_\textbf{k}^2+
(i\omega+\eta)^2[1+\chi(-\omega)]}}
 \fas e^{(i\omega+\eta)(t-t')}.
\end{eqnarray}
In equation (\ref{e128}) the relation
$\tilde{\chi}(i\omega)=\chi(-\omega)$ have been used where
$\tilde{\chi}(s)$ and $\chi(\omega)$ are the Laplace and Fourier
transforms of $\chi(t)$ respectively.

In equation (\ref{e128}) we have two situations: (i) The medium is
nondispersive and $\chi$ is a constant. In this case we use the
limit $t'\rightarrow 0$ for simplicity and solve the equation
(\ref{e128}) by using residue calculations and find

\begin{equation}\label{e129}
z_+(\omega ,t )=(1+n)\fas e^{(\frac{i\omega_k}{n}t)} +(1-n)\fas
e^{-(\frac{i\omega_k}{n}t)},
\end{equation}
where $n=\sqrt{\varepsilon}$ which coincide with the result of
\cite{9}.

 (ii) The medium is dissipative. In this case
 according to the Kramers$-$Kronig relations $\chi(\omega)$ is a
 complex function of $\omega$. Since $t-t'>0$, the
poles with positive imaginary part are important. Let there be N
poles with $\omega_i>0$ then

\begin{eqnarray}\label{e130}
{\lim\limits_{t'\to-\infty}}z_+(\omega _\textbf{k},t - t')
&=&\sum\limits_{n=1}^N{\lim
\limits_{t'\to-\infty}\alpha_n(\omega_k
)e^{i\omega_n(t-t')}},\nonumber\\&=&\sum\limits_{n=1}^N{\lim\limits_{t'\to-\infty}
[\alpha_n(\omega_k)e^{i\omega_{nr}(t-t')-\omega_{ni}t}]e^{\omega
_{ni}t'}},
\end{eqnarray}
where

\begin{equation}\label{e131}
\alpha_n(\omega_k)=\lim\limits_{\omega\to\omega_n}(\omega-\omega_n)
\{\frac{\omega[1-\chi(\omega)]-\omega_k}{\omega_k^2-\omega^2[1+\chi(-\omega)]}\}.
\end{equation}

From Eq.(\ref{e130}) it is clear that

\begin{equation}\label{e132}
\lim\limits_{t'\to-\infty}z_+(\omega_{\textbf{k}},t-t')\to0.
\end{equation}

The same procedure can be used to obtain the explicit form of the
second term of equation (\ref{e124}) and the final result is

\begin{eqnarray}\label{e133}
&&\xi(\omega_{\textbf{k}},\omega,t)= F(\omega)\nonumber
\\&\times &\lim\limits_{\eta\to0^+}\lim\limits_{t'\to-\infty}
\int\limits_{-\infty}^{+\infty} {\frac{d\omega'}{2\pi}
\frac{(i\omega+\eta)
e^{(i\omega+\eta)(t-t')}}{\{i(\omega+\omega')+\eta\}\{\omega_{\textbf{k}}^2-
(\omega-i\eta)^2[1+\chi(-\omega)]\}}}.\nonumber\\
\end{eqnarray}
In equation (\ref{e133}) there is one real pole and the other poles
have imaginary parts and tend to zero. Now by using the calculus of
residues we obtain

\begin{eqnarray}\label{e134}
\hat{\textbf{A}}(\textbf{r},t)&=&
II=\frac{1}{\epsilon_0}\sum\limits_{\lambda=1,2}{\int_0^\infty
{d\omega}\int\frac{d^3\textbf{k}}{(2\pi)^{\frac{3}{2}}}\fas
F(\omega) \frac{\omega}
{(\omega_{\textbf{k}}^2-\omega^2\varepsilon(\omega))}}
\nonumber\\&\times &{ (\hat{B}_\lambda(\textbf{k},\omega)e^{i(\omega
t-\textbf{k}\cdot\textbf{x})}+
H.c.})\fas\textbf{e}_{\lambda}(\textbf{k}),
\end{eqnarray}
where we have defined $\hat{B}_\lambda(\textbf{k},\omega)$ as

\begin{equation}\label{e135}
\lim\limits_{t'\to-\infty}\hat{B}_\lambda
(\textbf{k},\omega,t')e^{i\omega
t'}=\hat{B}_\lambda(\textbf{k},\omega).
\end{equation}
Using (\ref{e89}) to write $F(\omega)$ in terms of $Im \chi(\omega)$
equation (\ref{e134}) can be written as

\begin{eqnarray}\label{e136}
\hat{\textbf{A}}(\textbf{r},t)
 &=& \left({\frac{\hbar}{8\pi^4\varepsilon_0 }}
 \right)^{\frac{1}{2}} \int{d\omega
 \int{d^3{\textbf{k}}\sum\limits_{\lambda=1,2}
 {\omega \frac{\sqrt {Im\chi(\omega)}}{\textbf{k}^2 c^2
 -\omega^2\varepsilon(\omega)}}}}
 \nonumber \\
 &\times&(\hat{B}_\lambda(\textbf{k},\omega)e^{i(\omega t-\textbf{k}
 \cdot\textbf{r})} +
 H.c.)\dis\textbf{e}_\lambda(\textbf{k}).
 \end{eqnarray}
This equation is the same equation obtained in the phenomenological
and the damped polarization model \cite{3,6}.

By the same method the longitudinal part of electric field can be
obtained as
\begin{eqnarray}\label{e137}
\hat{ \textbf{E}}^\pa(\textbf{r},t)&=&\left({\frac{\hbar}{8\pi ^4
\varepsilon_0}}
  \right)^{\frac{1}{2}}
  \int_0^\infty d\omega\int d^3k\fas\frac{i\sqrt{Im\chi(\omega)}}
  {1+\chi (\omega)}\nonumber \\
 &\times& (\hat{B}_3(\textbf{k},\omega)e^{i(\omega
 t-\textbf{k}\cdot\textbf{r})}-h.c.)\fas\textbf{e}_3(\textbf{k}),
 \end{eqnarray}
which is $\frac{\partial \hat{\textbf{A}}^\parallel}{\partial t}$ in
the phenomenological model and also the longitudinal part of the
electric filed in the damped polarization model.

\section{Comparing the different methods}

In subsection 2.3 we showed that the phenomenological method can be
obtained from minimal coupling method and in section 3 we obtained
the minimal coupling Hamiltonian from the Lagrangian of damped
polarization method. In addition as can be seen from
(\ref{e55})-(\ref{e59}), (\ref{e73})-(\ref{e75}) and
(\ref{e96})-(\ref{e102}), these three methods lead to the same
results. Then in fact these three models are equivalent. In other
word they are different techniques for solving the equation of
motion of the same Lagrangian.

Although these three methods have the same results and they are
equivalent but each of them have its own advantage and disadvantage.

The merit of damped polarization method with respect to the other
methods is that it is based on a Lagrangian and then we can use the
standard canonical quantization method and the ETCR between
different operators can be deduced from its Lagrangian. But by
comparing the solution technique of three methods we see that the
solution technique of this model, which is based on Fano technique
and diagonalization of the Hamiltonian, is hard and lengthy. In
addition the extension of this model to nonisotropic and
nonhomogeneous polarizable and magnetizable medium has not been
down.

The preference of phenomenological method is that its way of solving
the equation is based on the Green function of the classical Maxwell
equations. This correlation function would be useful in calculating
the vacuum effect, like spontaneous emission, Casimir effect and Van
der Wales force \cite{16}$-$\cite{19}. In addition the extension of
this model to a nonhomogeneous and nonisotropic medium is easy and
only we should solve the classical Green function of the medium with
nonhomogeneous and nonisotropic susceptibility\cite{20}. Also the
extension of this model to a magnetizable medium is down by adding
new noise operators for the loss of magnetization \cite{21}. But the
disadvantage of this model is that it is not based on a Lagrangian
and the commutation relation between vector potential and electric
field is postulated and should be checked after quantization. In
addition in \cite{22} it is shown that in the minimal coupling
method the spontaneous emission can be solved easier than this
model.

The advantage of minimal coupling model is that in this model
Maxwell equations are obtained from Heisenberg equation. Also, as
mentioned in (\ref{e91}), the commutation relation between noise
operators are obtained from the Heisenberg equation which give a
better understanding of the nature of noise operators.

Another merit of this method is that it can be extended easily to a
nonhomogeneous, non isotropic and  nonlocal polarizable and
magnetizable medium \cite{9} and \cite{23}.

The disadvantage of this model is that since this model is not based
on a Lagrangian then the commutation between vector potential
$\hat{A}$ and the displacement vector $\hat{D}$ is postulated.

Therefore up to the range of applicability of these three models, we
could prove that these models are equivalent in a spatially
homogeneous and non magnetized medium.

\section{conclusion}

In this paper we have shown that the minimal coupling method is
equivalent to the Huttner-Barnet and phenomenological approaches up
to a canonical transformation in a nonmagnetic medium. The magnetic
properties of the medium are also included in the minimal coupling
method contrary to the other methods. So for a general comparison,
an extension of the Huttner-Barnet model to the case of a
magnetodielectric medium is needed which is under consideration.


\begin{thebibliography}{99}
\bibitem {1} S. M. Barnett, R. Matloob and R. Lodoun, J. Mod. Opt.
42, 1165 (1995)
\bibitem {2} R. J. Glauber and M. Lewnstein, Phys. Rev. A 43, 467 (1991)
\bibitem {3} B. Huttner and S. Barnett, Phys. Rev. A 46, 4306 (1992)
\bibitem {4} R. Matloob and R. Loudon, Phys. Rev. A 53, 4567 (1996)
\bibitem {5} R. Matloob and R. Loudon, Phys. Rev. A 52, 4623 (1995)
\bibitem {6} R. Matloob, Phys. Rev. A 60, 50 (1999)
\bibitem {7} F. Kheirandish and M. Amooshahi, Int. J. Theo. Phys.
45, No. 1 (2006)
\bibitem {8} F. Kheirandish and M. Amooshahi, Mod. Phys. Lett. A 29,
No 30 3025 (2005)
\bibitem {9} F. Kheirandish and M. Amooshahi, Phys. Rev. A 74,
042102 (2006)

\bibitem {11} V. Fano, Phys. Rev. 124, 1886 (1961)
\bibitem {12} S. M. Barnet and P. M. Radmor, Opt.Commun. 68, 304
(1998)
\bibitem {13} J. J. Hopffield, Phys. Rev. 112, 1555 (1958)
\bibitem {14}L. D. Landau and E. M. Lifshitz, \textit{Statistical Physics},
3rd ed. (Pergamon, Oxford, 1980), Part 1, Sec. 123
\bibitem {15}J. D. Jackson, classical Electrodynamics, 3rd ed. (Wiley, New York, 1999)
\bibitem {16}R. Matloob and H. Falinejad, phys. Rev. A 64, 042102 (2001)
\bibitem {17}S. Sheel, L. Knoll and D. G. Welsch, Phys. Rev. A
60,4094 (1999)
\bibitem {18}C. Rabba and D. G. Welcsh, Phys. Rev. A 73,063822 (2006)
\bibitem {19}S. Spagnolo, D. A. R. Dalvit and P. W. Milonni,  Phys.
Rev. A 75, 052117 (2007)
\bibitem {20}R. Matloob, Phys. Rev. A 71, 062105 (2005)
\bibitem {21}R. Matloob, Phys. Rev. A 77, 062103 (2005)
\bibitem {22} M. Amooshahi and F. Kheirandish, Phys. Rev. A 76, 062103 (2007)
\bibitem {23} F. Kheirandish and M. Amooshahi, arXiv:0705.3942, to
be published in Mod. Phys. Lett. A

\end{thebibliography}
\end{document}